\input{epsf} 
\newcommand{\et}{\hspace{-0.08in}{\bf .}\hspace{0.1in}}
\newcommand{\BOX}{\hbox {$\sqcap$ \kern -1em $\sqcup$}}
\newcommand{\qed}{\hskip 2em \hbox{\BOX} \vskip 2ex}

\renewcommand{\to}{\rightarrow}
\newcommand{\tensor}{\otimes}
\newcommand{\maps}{\colon}

\newcommand{\hf}{\frac{1}{2}}

\newcommand{\iso}{\cong}

\renewcommand{\H}{H}

\newcommand{\tr}{{\rm tr}}

\newcommand{\we}{\wedge}

\newcommand{\SU}{{\rm SU}}
\newcommand{\U}{{\rm U}}

\newcommand{\SO}{{\rm SO}}
\newcommand{\ISO}{{\rm ISO}}

\renewcommand{\c}{{\mathfrak c}}
\newcommand{\g}{{\mathfrak g}}
\newcommand{\h}{{\mathfrak h}}

\newcommand{\Bitors}{{\rm Bitors}}
\newcommand{\Der}{{\rm Der}}

\newcommand{\Aut}{{\rm Aut}}

\newcommand{\R}{{\mathbb R}}

\newtheorem{thm}{Theorem}    

\newtheorem{prop}[thm]{Proposition}
\newtheorem{defn}[thm]{Definition}
\newtheorem{example}[thm]{Example}

        \newcommand{\be}{\begin{equation}}
        \newcommand{\ee}{\end{equation}}
        \newcommand{\ba}{\begin{eqnarray}}
        \newcommand{\ea}{\end{eqnarray}}
        \newcommand{\ban}{\begin{eqnarray*}}
        \newcommand{\ean}{\end{eqnarray*}}
        \newcommand{\barr}{\begin{array}} 
        \newcommand{\earr}{\end{array}}

\documentclass{article}
\usepackage{amsfonts,amssymb}
\usepackage{diagram}
\usepackage[all]{xy}

      \title{Higher Yang--Mills Theory }
      \author{John C.\ Baez \\
      Department of Mathematics,  University of California\\ 
      Riverside, California 92521 \\
      USA \\
         \\
      email: baez@math.ucr.edu \\}
      \date{June 15, 2002}
     
\begin{document}
\bibliographystyle{plain}
\maketitle

\begin{abstract}

\noindent Electromagnetism can be generalized to Yang--Mills theory by
replacing the group $\U(1)$ by a nonabelian Lie group.  This raises the
question of whether one can similarly generalize 2-form electromagnetism
to a kind of `higher-dimensional Yang--Mills theory'.  It turns out that
to do this, one should replace the Lie group by a `Lie 2-group', which
is a category $C$ where the set of objects and the set of morphisms are
Lie groups, and the source, target, identity and composition maps are
homomorphisms.  We show that this is the same as a `Lie crossed module':
a pair of Lie groups $G,H$ with a homomorphism $t \maps H \to G$ and an
action of $G$ on $H$ satisfying two compatibility conditions.  Following
Breen and Messing's ideas on the geometry of nonabelian gerbes, one can
define `principal 2-bundles' for any Lie 2-group $C$ and do gauge theory
in this new context.  Here we only consider trivial 2-bundles, where a
connection consists of a $\g$-valued 1-form together with an $\h$-valued
2-form, and its curvature consists of a $\g$-valued 2-form together with
a $\h$-valued 3-form.  We generalize the Yang--Mills action for this
sort of connection, and use this to derive `higher Yang--Mills
equations'.  Finally, we show that in certain cases these equations
admit self-dual solutions in five dimensions.

\end{abstract}

\section{Introduction}

Describing electromagnetism in terms of the 1-form $A$ is very
natural, since to compute the action of a charged particle we merely
integrate this 1-form over the particle's path and add the result to the
action for a neutral particle moving along the same path.   A 2-form $B$
called the Kalb--Ramond field plays a similar role in the theory of
strings: we integrate this over the string worldsheet to get  a term in
the action.  Extending this analogy, one can write down equations for
`2-form electromagnetism' that are formally identical to Maxwell's
equations. Just as the electromagnetic field $F = dA$ automatically
satisfies $dF = 0$, so that the only nontrivial Maxwell equation is
\[     \ast d \ast F = J , \]
the field $G = dB$ automatically satisfies $dG = 0$, but one can 
impose an additional equation
\[     \ast d \ast G = K  \]
in which the current $K$ is now a 2-form.  Note that
just as the worldline of charged point particle defines a distributional
current 1-form, the worldsheet of a charged string defines 
a current 2-form, so the whole scheme is consistent.  

Since we can generalize Maxwell's equations to Yang--Mills theory by
replacing $A$ by a 1-form taking values in a nonabelian Lie algebra, it
is interesting to see if we can push the above analogy further and set
up some sort of `2-form Yang--Mills theory'.  There are indications from
string theory that something like this should exist \cite{Dijkgraaf},
but naive attempts to write down such a theory are already known
to fail \cite{Teitelboim}.  Fundamentally, the problem is that
there is no good way to define a notion of holonomy assigning elements
of a nonabelian group to the surfaces traced out by the motion of a
1-dimensional extended object.

To see the problem, first recall that it is very natural for holonomies
to assign elements of a nonabelian group to paths:
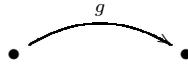
\[
\xymatrix{
\bullet\ar@/^1pc/[rr]^{g}="3"
     && \bullet  
}
\]
The reason is that composition of paths then corresponds to 
multiplication in the group:
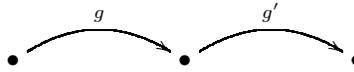
\[
\xymatrix{
\bullet\ar@/^1pc/[rr]^{g}="3"
     && \bullet \ar@/^1pc/[rr]^{g'}="3" &&\bullet 
}
\]
while reversing the direction of a path corresponds to taking
inverses:
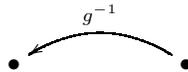
\[
\xymatrix{
\bullet
     && \bullet \ar@/_1pc/[ll]_{g^{-1}}="3"
}
\]
and the associative law makes the holonomy along a triple
composite unambiguous:
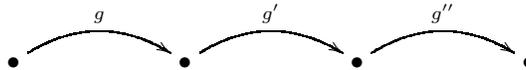
\[
\xymatrix{
\bullet\ar@/^1pc/[rr]^{g}="3"
     && \bullet \ar@/^1pc/[rr]^{g'}="3" 
     && \bullet \ar@/^1pc/[rr]^{g''}="3" 
     && \bullet 
} 
\]
In short, the geometry dictates the algebra.  

Now suppose we wish to do something similar for surfaces.  Naively
we might wish our holonomy to assign a group element to each
surface like this:
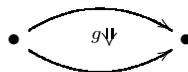
\[
\xymatrix{
\bullet \ar@/^1pc/[rr]^{}="3"
     \ar@/_1pc/[rr]_{}="4"
     && \bullet
\ar@{}"3";"4"^(.2){\,}="7"
     \ar@{}"3";"4"^(.7){\,}="8"
\ar@{=>}"7";"8"_{g}
}
\]
There are two obvious ways to compose surfaces of this sort, vertically:
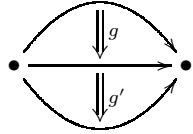
\[
 \xymatrix{
 \bullet
 \ar@/^2pc/[rr]^{\quad}_{}="0" 
 \ar[rr]^<<<<<<{\quad}_{}="1" 
\ar@/_2pc/[rr]_{\quad}_{}="2"
 &&  \bullet
 \ar@{=>}"0";"1"^{g}
\ar@{=>}"1";"2"^{g'}
}
\]
and horizontally:
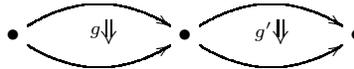
\[
\xymatrix{
\bullet \ar@/^1pc/[rr]^{\quad }="3"
     \ar@/_1pc/[rr]_{\quad }="4"
     && \bullet
\ar@{}"3";"4"^(.2){\,}="7"
     \ar@{}"3";"4"^(.7){\,}="8"
\ar@{=>}"7";"8"_{g}
\ar@/^1pc/[rr]^{\quad }="3"
     \ar@/_1pc/[rr]_{\quad }="4"
     && \bullet
\ar@{}"3";"4"^(.2){\,}="7"
     \ar@{}"3";"4"^(.7){\,}="8"
\ar@{=>}"7";"8"_{g'}
}
\]
Suppose that both of these correspond to multiplication
in the group $G$.  Then to obtain well-defined holonomy for this 
surface regardless of whether we do vertical or horizontal composition 
first:
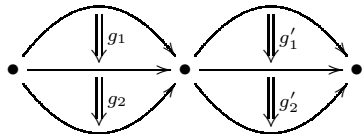
\[
 \xymatrix{
 \bullet
 \ar@/^2pc/[rr]^{\quad}_{}="0" 
 \ar[rr]^<<<<<<{\quad}_{}="1" 
\ar@/_2pc/[rr]_{\quad}_{}="2"
 &&  \bullet
 \ar@{=>}"0";"1"^{g_1}
\ar@{=>}"1";"2"^{g_2}
 \ar@/^2pc/[rr]^{\quad}_{}="0" 
 \ar[rr]^<<<<<<{\quad}_{}="1" 
\ar@/_2pc/[rr]_{\quad}_{}="2"
 &&  \bullet
 \ar@{=>}"0";"1"^{g'_1}
\ar@{=>}"1";"2"^{g'_2}
}  
\]
we must have 
\[      (g_1 g_2)(g_1' g_2') = (g_1 g_1')(g_2 g_2') . \]
This forces $G$ to be abelian!

In fact, this argument goes back to a classic paper by 
Eckmann and Hilton \cite{EH}.  Moreover, they showed
that even if we allow $G$ to be equipped with two products,
say $g \circ g'$ for vertical composition and $gg'$ for
horizontal, so long as both products share the
same unit and satisfy this `exchange law':
\[  (g_1 \circ g_1')(g_2 \circ g_2') =
      (g_1 g_2) \circ (g_1' g_2')  \]
then in fact they must agree --- so by the previous argument,
both are abelian.  The proof is very easy:
\[  gg' = (g \circ 1)(1 \circ g') = (g 1) \circ (1 g') =  g \circ g'. \]

\eject

The way around this roadblock is to consider a sort of connection
that allows us to define holonomies {\it both for paths and for surfaces}.
Assume that for each path we have a holonomy taking values in some
group $G$:
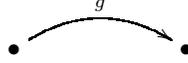
\[
\xymatrix{
\bullet\ar@/^1pc/[rr]^{g}="3"
     && \bullet  
}
\]
where composition of paths corresponds to multiplication in $G$.
Assume also that for each 1-parameter family of paths with 
fixed endpoints we have a holonomy taking values in some other 
group $H$:
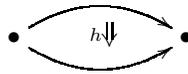
\[
\xymatrix{
\bullet \ar@/^1pc/[rr]^{\quad}="3"
     \ar@/_1pc/[rr]_{\quad}="4"
     && \bullet
\ar@{}"3";"4"^(.2){\,}="7"
     \ar@{}"3";"4"^(.7){\,}="8"
\ar@{=>}"7";"8"_{h}
}
\]
where vertical composition corresponds to multiplication in
$H$:
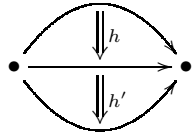
\[
 \xymatrix{
 \bullet
 \ar@/^2pc/[rr]^{\quad}_{}="0" 
 \ar[rr]^<<<<<<{\quad}_{}="1" 
\ar@/_2pc/[rr]_{\quad}_{}="2"
 &&  \bullet
 \ar@{=>}"0";"1"^{h}
\ar@{=>}"1";"2"^{h'}
}
\]

Next, assume that we can parallel transport an element $g \in G$
along a 1-parameter family of paths to get a new element $g' \in G$:
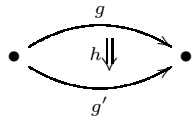
\[
\xymatrix{
\bullet \ar@/^1pc/[rr]^{g}="3"
     \ar@/_1pc/[rr]_{g'}="4"
     && \bullet
\ar@{}"3";"4"^(.2){\,}="7"
     \ar@{}"3";"4"^(.7){\,}="8"
\ar@{=>}"7";"8"_{h}
}
\]
Here a caveat is in order: if $g$ is the holonomy along the initial
path, we do not assume that $g'$ is the holonomy along the final path.
To require this would impose a kind of `flatness' condition on our
connection.  In the formalism we are heading toward, there will be a
$\g$-valued curvature 2-form $F$ that measures the deviation from this
sort of flatness.  There will also be an $\h$-valued 3-form that
measures how the $H$-valued holonomy along a 1-parameter family of paths
changes as we vary this family while keeping its initial and final paths
fixed.  

Now, the picture above suggests that we should think of $h$ as a kind of
`arrow' or `morphism' going from $g$ to $g'$, and use category theory to
formalize the situation.  However, in category theory, when a morphism
goes from an object $x$ to an object $y$, we think of the morphism as
determining both its source $x$ and its target $y$.  The group
element $h$ does not determine $g$ or $g'$.  However, the pair $(h,g)$
does.  Thus it is useful to create a category $C$ where the set $C_0$ of
objects is just $G$, while the set $C_1$ of morphisms consists
of pairs $f = (h,g) \in H \times G$.  
Switching our notation to reflect this, we rewrite
the above picture as
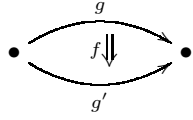
\[
\xymatrix{
\bullet \ar@/^1pc/[rr]^{g}="3"
     \ar@/_1pc/[rr]_{g'}="4"
     && \bullet
\ar@{}"3";"4"^(.2){\,}="7"
     \ar@{}"3";"4"^(.7){\,}="8"
\ar@{=>}"7";"8"_{f}
}
\]
and write $f \maps g \to g'$ for short.  We have source and target 
maps 
\[  s,t \maps C_1 \to C_0\]
with $s(f) = g$ and $t(f) = g'$.

In this new notation we can vertically compose $f \maps g \to g'$ and
$f' \maps g \to g'$ to get $f \circ f' \maps g \to g''$, as follows:
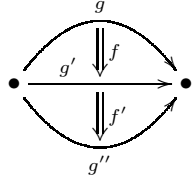
\[
 \xymatrix{
 \bullet
 \ar@/^2pc/[rr]^{g}_{}="0" 
 \ar[rr]^<<<<<<{g'}_{}="1" 
\ar@/_2pc/[rr]_{g''}_{}="2"
 &&  \bullet
 \ar@{=>}"0";"1"^{f}
\ar@{=>}"1";"2"^{f'}
}
\]
This is just composition of morphisms in the category $C$.  However, we
can also horizontally compose $f_1 \maps g_1 \to g_1'$ and $f_2 \maps
g_2 \to g_2'$ to get $f_1f_2 \maps g_1g_2 \to g_1'g_2'$, as follows:
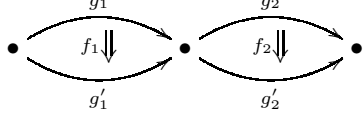
\[
\xymatrix{
\bullet \ar@/^1pc/[rr]^{g_1}="3"
     \ar@/_1pc/[rr]_{g_1'}="4"
     && \bullet
\ar@{}"3";"4"^(.2){\,}="7"
     \ar@{}"3";"4"^(.7){\,}="8"
\ar@{=>}"7";"8"_{f_1}
\ar@/^1pc/[rr]^{g_2}="3"
     \ar@/_1pc/[rr]_{g_2'}="4"
     && \bullet
\ar@{}"3";"4"^(.2){\,}="7"
     \ar@{}"3";"4"^(.7){\,}="8"
\ar@{=>}"7";"8"_{f_2}
}
\]
We assume this operation makes $C_1$ into a group with the pair
$(1,1) \in H \times G$ as its multiplicative unit.

The good news is that now we can assume an exchange law saying this
holonomy is well-defined:
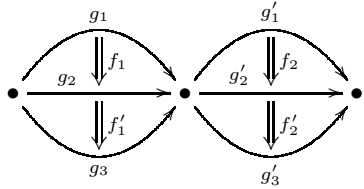
\[
 \xymatrix{
 \bullet
 \ar@/^2pc/[rr]^{g_1}_{}="0" 
 \ar[rr]^<<<<<<{g_2}_{}="1" 
\ar@/_2pc/[rr]_{g_3}_{}="2"
 &&  \bullet
 \ar@{=>}"0";"1"^{f_1}
\ar@{=>}"1";"2"^{f_1'}
 \ar@/^2pc/[rr]^{g_1'}_{}="0" 
 \ar[rr]^<<<<<<{g_2'}_{}="1" 
\ar@/_2pc/[rr]_{g_3'}_{}="2"
 &&  \bullet
 \ar@{=>}"0";"1"^{f_2}
\ar@{=>}"1";"2"^{f_2'}
}
\]
namely:
\[ (f_1 \circ f_1')(f_2 \circ f_2') =  (f_1 f_2) \circ (f_1' f_2') ,\]
without forcing either $C_0$ or $C_1$ to be abelian!   Instead,
$C_1$ is forced to be a semidirect product of the groups $G$ and
$H$.

The structure we are rather roughly describing here is in fact already
known to mathematicians under the name of a `categorical group'
\cite{FB,MacLane}.  The reason is that $C$ turns out to be a category
whose set of objects $C_0$ is a group, whose set of morphisms $C_1$ is a
group, and where all the usual category operations are group
homomorphisms.  To keep the terminology succinct and to hint at
generalizations to still higher-dimensional holonomies, we prefer to
call this sort of structure a `2-group'.  Moreover, we shall focus most
of our attention on `Lie 2-groups', where $C_0$ and $C_1$ are Lie groups
and all the operations are smooth.

As we shall see, a Lie 2-group $C$ amounts to the same thing as a `Lie
crossed module': a pair of Lie groups $G$ and $H$ together with a
homomorphism $t \maps H \to G$ and an action $\alpha$ of $G$ on $H$
satisfying a couple of compatibility conditions.  The idea is to let $G
= C_0$, let $H$ be the subgroup of $C_1$ consisting of morphisms
with source equal to $1$, and let $t$ be the map sending each such
morphism to its target.  The action $\alpha$ is defined by letting
$\alpha(g)f$ be this horizontal composite:
\[
\xymatrix{
\bullet \ar@/^1pc/[rr]^{g}="3"
     \ar@/_1pc/[rr]_{g}="4"
     && \bullet
\ar@{}"3";"4"^(.2){\,}="7"
     \ar@{}"3";"4"^(.7){\,}="8"
\ar@{=>}"7";"8"_{1_g}
\ar@/^1pc/[rr]^{1}="3"
     \ar@/_1pc/[rr]_{t(f)}="4"
     && \bullet
\ar@{}"3";"4"^(.2){\,}="7"
     \ar@{}"3";"4"^(.7){\,}="8"
\ar@{=>}"7";"8"_{f}
\ar@/^1pc/[rr]^{g^{-1}}="3"
     \ar@/_1pc/[rr]_{g^{-1}}="4"
     && \bullet
\ar@{}"3";"4"^(.2){\,}="7"
     \ar@{}"3";"4"^(.7){\,}="8"
\ar@{=>}"7";"8"_{1_{g^{-1}}}
}
\]

It appears that one can develop a full-fledged theory of bundles,
connections, curvature, the Yang--Mills equations, and so on with a Lie
2-group taking the place of a Lie group.  So far most work has focused
on the special case when $G$ is trivial and $H$ is abelian
\cite{Brylinski,CJM,CMR,Hitchin,Keurentjes,MP}, but here we really want
$H$ to be nonabelian.  Some important progress in this direction can be
found in Breen and Messing's paper on the differential geometry of
nonabelian gerbes \cite{BM}, and also Attal's work on a combinatorial
version of this theory \cite{Attal}.  While they use different
terminology, their work basically develops the theory of connections and
curvature for Lie 2-groups where $H$ is an arbitrary Lie group, $G =
\Aut(H)$, $t$ sends each element of $H$ to the corresponding inner
automorphism, and the action of $G$ on $H$ is the obvious one.  We call
this sort of Lie 2-group the `automorphism 2-group' of $H$.  Luckily, it
is easy to extrapolate the whole theory from this case.

In particular, for any Lie 2-group $C$ one can define the notion of a
`principal 2-bundle' having $C$ as its gauge 2-group, and then
define connections and curvature for these principal 2-bundles.
However, to keep things simple, we do this here only for {\it trivial}
principal 2-bundles.  This renders the whole apparatus of 2-bundles
unnecessary, since a connection then amounts to just a
$\g$-valued 1-form $A$ and an $\h$-valued 2-form $B$ on the base
manifold $M$.  The curvature consists of a $\g$-valued 2-form:
\[      F = dA + \hf A \we A - \underline{B}  \]
together with a $\h$-valued 3-form:
\[      G = dB + \underline{A \we B}  \]
where the $\g$-valued 2-form $\underline{B}$ is obtained from $B$ using
the map $dt \maps \h \to \g$, and the $\h$-valued 3-form $\underline{A
\we B}$ is defined using the action $d\alpha$ of $\g$ on $\h$.  The
suprising new $\underline{B}$ term in the curvature 2-form $F$
comes from the fact that we use 2-dimensional parallel transport
to compare holonomies along nearby paths, instead of merely taking their
difference.

The curvature automatically satisfies these `higher Bianchi identities':
\[
\begin{array}{lcl} 
       dF + A \we F &=& -\underline{G},  \\
       dG + \underline{A \we G} &=& \underline{F \we B} .  \\
\end{array}
\]
Moreover, we can develop a generalization of Yang--Mills theory
whenever the base manifold $M$ is oriented and
equipped with a metric, and $\g$ and $\h$ are equipped with invariant
inner products.  To do this, we start with a generalization of
the Yang--Mills action:
\[   S(A,B) = \int_M \tr(F \we \ast F) + \tr(G \we \ast G)  .\]
Extremizing this action we obtain the `higher Yang--Mills equations':
\[
\begin{array}{lcl}
d \ast F + A \we \ast F &=& \underline{\ast G \we B} .   \\
d \ast G + \underline{A \we \ast G} &=& - \ast \underline{F}  
\end{array}  
\]
Here the underlines again refer to certain natural maps involving $\g$
and $\h$.  In the special case when $H$ is trivial, these equations
reduce to the ordinary Yang--Mills equations with gauge group $G$. 
When $G$ is trivial and $H = \U(1)$, the higher Yang--Mills equations 
reduce to those of 2-form electromagnetism.  

In short, by replacing the concept of `Lie group' by the concept
of `Lie 2-group', we arrive at equations that constitute a simultaneous
generalization of Yang--Mills theory and 2-form electromagnetism.  It
remains to see how useful these equations are, either in physics or in
pure mathematics.  However, we shall show that under certain conditions on
the Lie 2-group $C$, they admit `self-dual' solutions in 5-dimensional
spacetime.  Since the self-dual solutions of the ordinary Yang--Mills
equations are important in 4-dimensional topology, it is tempting to
hope that the higher Yang--Mills equations will play an interesting role
in topology one dimension up.

The plan of the paper is as follows.  In Section \ref{Lie2group} we
carefully define Lie 2-groups, show they are the same as Lie crossed
modules, and give a number of examples.  In Section \ref{Lie2group} we
define Lie 2-algebras, show they are the same as `differential crossed
modules', and show how to get a Lie 2-algebra from a Lie 2-group.  In
Section \ref{connections} we define connections and curvature for
trivial 2-bundles, and prove the higher Bianchi identities in this case.
Finally, in Section \ref{ymeqs} we derive the higher Yang--Mills
equations from an action principle, and show they have self-dual
solutions in 5 dimensions.  There is much we leave undone: for example,
we do not develop the general theory of 2-bundles, nor do we work out 
the theory of holonomies for the connections we consider, even though 
it serves as a vital motivation for the concept of Lie 2-group.  

\section{Lie 2-Groups} \label{Lie2group}

The concept of `Lie 2-group' is a kind of blend of the concepts of Lie
group and category.   A small category $C$ has a set $C_0$ of objects,
a set $C_1$ of morphisms, functions 
\[  s,t \maps C_1 \to C_0 \]
assigning to each morphism $f \maps x \to y$ its source $x$
and target $y$, a function
\[  i \maps C_0 \to C_1 \]
assigning to each object its identity morphism, and finally, a function 
\[  \circ \maps C_1 \times_{C_0} C_1 \to C_1  \]
describing composition of morphisms, where 
\[  C_1 \times_{C_0} C_1 = 
\{(f,g) \in C_1 \times C_1 \, \colon \; t(f) = s(g) \}   \]
is the set of composable pairs of morphisms.
If we now take the words `set' and `function' and replace them by `Lie
group' and `homomorphism', we get the definition of a Lie 2-group.
Here and in all that follows, we require that homomorphisms between
Lie groups be smooth:

\begin{defn}  \et
A {\bf Lie 2-group} is a category $C$ where the set $C_0$ of objects and
the set $C_1$ of morphisms are Lie groups, the functions $s,t \maps C_1
\to C_0$, $i \maps C_0 \to C_1$ are homomorphisms, $C_1
\times_{C_0} C_1$ is a Lie subgroup of $C_1 \times C_1$, and $\circ
\maps  C_1 \times_{C_0} C_1 \to C_1$ is a homomorphism.
\end{defn}

\noindent 
The fact that composition is a homomorphism implies the
exchange law
\[ (f_1 \circ f_1')(f_2 \circ f_2') =  (f_1 f_2) \circ (f_1' f_2') \]
whenever we have a situation of this sort:
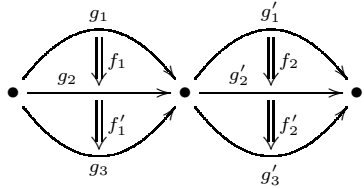
\[
 \xymatrix{
 \bullet
 \ar@/^2pc/[rr]^{g_1}_{}="0" 
 \ar[rr]^<<<<<<{g_2}_{}="1" 
\ar@/_2pc/[rr]_{g_3}_{}="2"
 &&  \bullet
 \ar@{=>}"0";"1"^{f_1}
\ar@{=>}"1";"2"^{f_1'}
 \ar@/^2pc/[rr]^{g_1'}_{}="0" 
 \ar[rr]^<<<<<<{g_2'}_{}="1" 
\ar@/_2pc/[rr]_{g_3'}_{}="2"
 &&  \bullet
 \ar@{=>}"0";"1"^{f_2}
\ar@{=>}"1";"2"^{f_2'}
}
\]

A homomorphism between Lie 2-groups is a kind of blend
of the concepts of homomorphism between Lie groups and functor
between categories:

\begin{defn} \label{2homo1} \et
Given Lie 2-groups $C$ and $C'$, a {\bf homomorphism} $F \maps C \to C'$
is a functor which acts on objects and morphisms to give homomorphisms
$F_0 \maps C_0 \to C_0'$ and $F_1 \maps C_1 \to C_1'$, respectively.
\end{defn}

A Lie 2-group with only identity morphisms is the same thing as a 
Lie group.  To get more interesting examples it is handy to think of a
Lie 2-group as special sort of crossed module.
To do this, start with a Lie 2-group $C$ 
and form the pair of Lie groups 
\[   G = C_0  , \qquad H = \ker s \subseteq C_1.  \]
The target map restricts to a homomorphism 
\[   t \maps H  \to G . \]
Besides the usual action of $G$ on itself by conjugation, 
there is also an action of $G$ on $H$,
\[    \alpha \maps G \to \Aut(H) ,\]
given by
\[   \alpha(g)(h)   =  i(g)\, h \, i(g)^{-1} . \]
The target map is equivariant with respect to this 
action:
\[   t(\alpha(g)(h) = g \, t(h)\, g^{-1}. \]
and we also have
\[    \alpha(t(h))(h') = hh'h^{-1}  \]
for all $h,h' \in H$.  A setup like this with groups rather than Lie
groups is called a `crossed module', so here we are getting a `Lie
crossed module':

\begin{defn} \et
A {\bf Lie crossed module} is a quadruple $(G,H,t,\alpha)$
consisting of Lie groups $G$ and $H$, a homomorphism $t \maps H \to G$,
and an action of $G$ on $H$ (that is, a homomorphism 
$\alpha \maps G \to \Aut(H)$) satisfying
\[   t(\alpha(g)(h)) = g \, t(h)\, g^{-1} \]
and 
\[    \alpha(t(h))(h') = hh'h^{-1}  \]
for all $g \in G$ and $h,h' \in H$.
\end{defn}

\noindent This definition becomes a bit more memorable if
allow ourselves to write $\alpha(g)(h)$ as $ghg^{-1}$; then the
equations above become
\[   t(ghg^{-1}) = g \, t(h)\, g^{-1} \]
and 
\[    t(h) \, h' \, t(h)^{-1} = hh'h^{-1} . \]

\begin{defn} \et
A {\bf homomorphism} from the Lie crossed module $(G,H,t,\alpha)$ to the
Lie crossed module $(G',H',t',\alpha')$ is a commutative square of
homomorphisms 
\[
\begin{diagram}[H]
\node{H} \arrow{e,t}{\tilde f} \arrow{s,l}{t} \node{H'} \arrow{s,r}{t'} \\
\node{G} \arrow{e,t}{f} \node{G'} 
\end{diagram}
\]
such that
\[   f(\alpha(g)(h)) = \alpha'({\tilde f}(g)) (f(h)) \]
for all $g \in G$, $h \in H$.
\end{defn}

\begin{prop} \label{liecrossedmodule} \et
The category of Lie 2-groups is equivalent to the category of Lie
crossed modules. \end{prop}

Proof --- This follows easily from the well-known equivalence between
crossed modules and 2-groups, also known as categorical groups \cite{BS}
or sometimes gr-categories \cite{Breen2}.  A nice exposition of this can
be found in a paper by Forrester-Barker \cite{FB}, but for the
convenience of the reader we sketch how to recover a Lie 2-group from a
Lie crossed module.

Suppose we have a Lie crossed module $(G,H,t,\alpha)$.  Let 
\[   C_0 = G, \qquad C_1 = H \rtimes G  \]
where the semidirect product is formed using the action of $G$
on $H$, so that multiplication in $C_1$ is given by
\[    (h,g) (h',g') = (h \alpha(g)(h'), gg') .\]
We make this into a Lie 2-group where the source
and target maps $s,t \maps C_1 \to C_0$ are given by
\[   s(h,g) = g , \qquad   t(h,g) = t(h) g  ,\]
the map $i \maps C_0 \to C_1$ is given by
\[  i(g) = (1,g), \]
and the composite of the morphisms
\[  (h,g) \maps g  \to g'  , \qquad \quad
  (h',g') \maps g' \to g'' \]
is 
\[   (hh',g) \maps g \to g'' .\]
One can check that this construction indeed gives a Lie 2-group, and
that together with the previous construction it sets up an equivalence
between the categories of Lie 2-groups and Lie crossed modules.  \qed

Crossed modules are important in homotopy theory \cite{Brown},
and the reader who is fonder of crossed modules than categories 
is free to think of Lie 2-groups as an idiosyncratic way of talking
about Lie crossed modules.  Both perspectives are useful, but one
advantage of Lie crossed modules is that they allow us to
quickly describe some examples:

\begin{example} \label{crossed.1} \et {\rm 
Given any Lie group $G$, abelian Lie group $H$, and
homomorphism $\alpha \maps G \to \Aut(H)$, 
there is a Lie crossed module with $t \maps
G \to H$ the trivial homomorphism and $G$ acting on $H$ via $\rho$. 
Because $t$ is trivial, the corresponding Lie 2-group $C$ 
is `skeletal', meaning that any two isomorphic objects are equal.  
It is easy to see that conversely, all skeletal Lie 2-groups are of 
this form.
}\end{example}

\begin{example} \label{crossed.2} \et {\rm 
Given any Lie group $G$, we can form a Lie crossed module as in
Example \ref{crossed.1} by taking $H = \g$, thought of as an abelian 
Lie group, and letting $\alpha$ be the adjoint representation of $G$
on $\g$.  If $C$ is the corresponding Lie 2-group, we have
\[    C_1 = \g \rtimes G \iso TG \]
where $TG$ is the tangent bundle of $G$, which becomes a Lie
group with product 
\[    dm \maps TG \times TG \to TG , \]
obtained by differentiating the product
\[    m \maps G \times G \to G .\]
We call $C$ the {\bf tangent 2-group} of $G$.
}\end{example}

\begin{example} \label{crossed.3} \et {\rm
Similarly, given any Lie group $G$, we can form a Lie crossed module 
as in Example \ref{crossed.1} by letting $\alpha$ be the coadjoint 
representation on $H = \g^*$.
If $C$ is the corresponding Lie 2-group, we have
\[    C_1 = \g^* \rtimes G \iso T^*G \]
where $T^*G$ is the cotangent bundle of $G$, 
and we call $C$ the {\bf cotangent 2-group} of $G$.
}\end{example}

\begin{example} \label{crossed.4} \et {\rm 
If $G$ is the Lorentz group $\SO(n,1)$, 
we can form a Lie crossed module as in Example 
\ref{crossed.1} by letting $\alpha$ be the defining representation 
of $\SO(n,1)$ on $H = \R^{n+1}$. 
If $C$ is the corresponding Lie 2-group, we have
\[    C_1 = \R^{n+1} \rtimes \SO(n,1) \iso \ISO(n,1) \]
where $\ISO(n,1)$ is the Poincar\'e group.  We call $C$ the {\bf
Poincar\'e 2-group}.
}\end{example}

\begin{example} \label{crossed.5} \et {\rm
Given any Lie group $H$, there is a Lie
crossed module with $G = \Aut(H)$, $t \maps H \to G$ the homomorphism
assigning to each element of $H$ the corresponding inner automorphism,
and the obvious action of $G$ as automorphisms of $H$.  We call the
corresponding Lie 2-group $C$ the {\bf automorphism 2-group} of $H$.
}\end{example}

\begin{example} \label{crossed.6} \et {\rm
If we take $H = \SU(2)$ and form the automorphism 2-group of $H$, we get
a Lie 2-group with $G = \SO(3)$.  Since $\SU(2)$ can be thought of as
the unit quaternions, this example is closely related to a larger Lie
2-group implicit in Thompson's work on `quaternionic gerbes'
\cite{Thompson}.  This larger 2-group is just the automorphism 2-group
of the multiplicative group of nonzero quaternions.  
}\end{example}

Example \ref{crossed.5} is important in the theory of nonabelian
gerbes.  Any Lie 2-group $C$ becomes a strict monoidal category if
we define a tensor product
\[  \tensor \maps C \times C \to C \]
by
\[   g \tensor g'   = gg' \]
for all objects $g,g' \in C_0$, and
\[   f \tensor f' = ff' \]
for all morphisms $f,f' \in C_1$, where $ff'$ is defined using the
product in the group $C_1$, not composition of morphisms.  
When $C$ is the automorphism 2-group of $H$, this monoidal category 
is equivalent to the monoidal category of `$H$-bitorsors'.  An {\bf
$H$-bitorsor} $X$ is a manifold with commuting left and right actions of
$H$, both of which are free and transitive.  A morphism between
$H$-bitorsors $f \maps X \to Y$ is a smooth map which is equivariant
with respect to both these actions.  The `tensor product' of
$H$-bitorsors $X$ and $Y$ is defined to be the space
\[   X \tensor Y =    X \times Y/((xh,y) \sim (x,hy)) , \]
which inherits a left $H$-action from $X$ and a right $H$-action from
$Y$.  Accompanying this there is an obvious notion of the tensor product
of morphisms between bitorsors, making $H$-bitorsors into a weak
monoidal category $\H$-$\Bitors$.   Every $H$-bitorsor is isomorphic
to $H$ with its usual right action on itself but with the left action
twisted by an automorphism of $H$.  Using this fact, it is not hard to 
check that $\H$-$\Bitors$ is equivalent to the automorphism 2-group of $H$.

One can describe $H$-gerbes in terms of $H$-bitorsors
\cite{Attal,Breen,BM}.  Alternatively, one can construct an $H$-gerbe by
taking the stack of sections of a `principal 2-bundle' where the `gauge
2-group' is the automorphism 2-group of $H$.  By the above remarks these
viewpoints are not really that different, although the latter one seems
not to have been developed in the literature.

Finally, for the reader familiar with categorification \cite{BD}, it is
worth mentioning that what we are calling Lie 2-groups should really be
called `strict' Lie 2-groups.  Ultimately more interesting will be the
more general `weak' ones, which will be weak rather than strict monoidal
categories, and where the inverse for an object $g$ satisfies the laws
$gg^{-1} = g^{-1}g = 1$ only up to isomorphism.  Our goal here is not to
develop higher Yang--Mills theory in its full generality, but merely to
show that it exists.

\section{Lie 2-Algebras} \label{Lie2algebra}

Just as a Lie 2-group is a category built from Lie groups,
a Lie 2-algebra is a category built from Lie algebras:

\begin{defn} \label{2liealgebra}\et  
A {\bf Lie 2-algebra} is a category $\c$ where the set $\c_0$ of objects
and the set $\c_1$ of morphisms are Lie algebras, and the functions $s,t
\maps \c_1 \to \c_0$, $i \maps \c_0 \to \c_1$ and $\circ \maps \c_1
\times_{\c_0} \c_1 \to \c_1$ are Lie algebra homomorphisms.
\end{defn}

\begin{defn} \label{2liehomo} \et
Given Lie 2-algebras $\c$ and $\c'$,
a {\bf homomorphism} $F \maps \c \to \c'$ is a functor which
acts on objects and morphisms to give Lie algebra
homomorphisms $F_0 \maps \c_0 \to \c_0'$ and $F_1 \maps \c_1 \to \c_1'$,
respectively.  
\end{defn}

Just as Lie groups have Lie algebras, Lie 2-groups have Lie 2-algebras:

\begin{prop} \label{lie.2-algebra.1} \et 
Any Lie 2-group $C$ has a Lie 2-algebra $\c$ for
which $\c_0$ is the Lie algebra of the Lie group of objects $C_0$,
$\c_1$ is the Lie algebra of the Lie group of morphisms $C_1$, and
the maps $s,t \maps \c_1 \to \c_0$, $i \maps \c_0 \to \c_1$
and $\circ \maps \c_1 \times_{\c_0} \c_1 \to \c_1$ are the differentials
of those for $C$.  Similarly, any homomorphism of Lie 2-groups
$F \maps C \to C'$ gives rise to a homomorphism of Lie 2-algebras,
which we denote by $dF \maps \c \to \c'$.  We obtain a functor
from the category of Lie 2-groups to the category of Lie 2-algebras.
\end{prop}

Proof --- The proof is straightforward. \qed

The equivalence between Lie 2-groups and Lie 
crossed modules has an analogue for Lie 2-algebras.  In this
analogue, the replacement for the Lie group $\Aut(H)$ is the 
Lie algebra $\Der(\h)$ consisting of all `derivations' of 
$\h$, that is, linear maps $f \maps \h \to \h$ such that 
\[          f([y,y']) = [f(y),y'] + [y,f(y')]  .\]

\begin{defn} \et
A {\bf differential crossed module} is a quadruple $(\g,\h,t,\alpha)$
consisting of Lie algebras $\g,\h$, a homomorphism $t \maps \h \to \g$,
and an action $\alpha$ of $\g$ as derivations of $\h$ 
(that is, a homomorphism $\alpha \maps \g \to \Der(\h)$) satisfying
\[   t(\alpha(x)(y)) = [x, t(y)] \]
and 
\[   \alpha(t(y)) (y') = [y, y'] \]
for all $x \in \g$ and $y,y' \in \h$.
\end{defn}

\noindent 
This definition becomes easier to remember if we allow ourselves to
write $\alpha(x)(y)$ as $[x,y]$.  Then the fact that $\alpha$ is an
action of $\g$ as derivations of $\h$ simply means that $[x,y]$ is
linear in each argument and the following `Jacobi identities' hold:
\begin{equation} \label{jacobi1}
         [[x,x'],y] = [x,[x',y]] - [x',[x,y]] ,   
\end{equation}
\begin{equation} \label{jacobi2}
         [x,[y,y']] = [[x,y],y'] - [[x,y'],y] 
\end{equation}
for all $x,x' \in \g$ and $y,y' \in \h$.
Furthermore, the two equations in the above definition become
\begin{equation} \label{cross1}
  t([x,y]) = [x, t(y)] 
\end{equation}
and 
\begin{equation} \label{cross2}
   [t(y), y'] = [y, y'] .
\end{equation}

\begin{prop} \label{lie.2-algebra.2}
\et The category of Lie 2-algebras is equivalent to the
category of differential crossed modules.  
\end{prop}

Proof --- The proof is just like that of Proposition \ref{liecrossedmodule}.
\qed

Any Lie crossed module gives rise to a differential crossed module; this
is just a less elegant way of saying that any Lie 2-group has a Lie
2-algebra:

\begin{prop} \label{diff.crossed.module} \et Any Lie crossed module
$(G,H,t,\alpha)$ has a differential crossed module $(\g,\h,dt,d\alpha)$
where $\g$ is the Lie algebra of $G$, $\h$ is the Lie algebra of $H$,
$dt \maps \h \to \g$ is the differential of $t \maps H \to G$, and the
action of $\g$ as derivations of $\h$ is the differential $d\alpha \maps
\g \to \Der(\h)$ of the homomorphism $\alpha \maps G \to \Aut(H)$.  We
obtain a functor from the category of Lie crossed modules to the
category of differential crossed modules.  \end{prop}

Proof --- This can easily be seen directly, but it also follows
from combining Propositions \ref{liecrossedmodule}, 
\ref{lie.2-algebra.1}, and \ref{lie.2-algebra.2}. \qed

\section{Connections and Curvature} \label{connections}

The concepts of connections and curvature can be generalized by
replacing Lie groups by Lie 2-groups and Lie algebras by Lie 2-algebras.
In particular, Breen and Messing \cite{BM} have defined these notions in
the special case when the `gauge 2-group' $C$ is of the special sort
described in Example \ref{crossed.5}: namely, the automorphism 2-group
of a Lie group.  Given this, generalizing to arbitrary Lie 2-groups is
not very hard.  However, to bring out the basic ideas in their simplest
form, we do this here only for `trivial principal 2-bundles', where
connections and curvature can be thought of as Lie-algebra-valued
differential forms:

\begin{defn} \et 
Suppose $C$ is a Lie 2-group and $(G,H,t,\alpha)$ is the corresponding
Lie crossed module.  A {\bf connection} on the manifold $M$ is a pair
$(A,B)$ consisting of a $\g$-valued 1-form $A$ on $M$ together with a
$\h$-valued 2-form $B$.  
\end{defn}

Now suppose that $C$ is Lie 2-group, $(G,H,t,\alpha)$ the corresponding
Lie crossed module, and $(\g,\h,dt,d\alpha)$ the differential crossed
module obtained via Proposition \ref{diff.crossed.module}.   

\begin{defn} \et Suppose $(A,B)$ is a connection on $M$.  Then the
{\bf curvature} of this connection is the pair $(F,G)$ where $F$ is the
$\g$-valued 2-form
\[      F = dA + \hf A \we A - \underline{B}  \]
and $G$ is the $\h$-valued 3-form
\[      G = dB + \underline{A \we B}  .\]
\end{defn}

Here we define the wedge product of $\g$-valued differential
forms by
\[    (x \tensor \mu) \we (x' \tensor \mu') = [x,x'] \tensor (\mu \we \mu') \]
where $x,x' \in \g$ and $\mu,\mu'$ are diferential forms; this
extends by linearity in each argument to all $\g$-valued differential forms.
Similarly, given a $\g$-valued differential form $X = x \tensor \mu$
and an $\h$-valued differential form $Y = y \tensor \nu$, we define
\[        \underline{X \we Y} = [x,y] \tensor (\mu \we \nu) \]
where for simplicity of notation we write $[x,y]$ as an 
abbrevation for $d\alpha(x)(y)$.  
Also, given an $\h$-valued differential form $Y = y \tensor \nu$, we
define
\[        \underline{Y} = dt(y) \tensor \nu \]
and extend this to arbitrary $\h$-valued differential forms by linearity.

We can streamline the definition of the curvature 3-form $G$ and the
statement of the Bianchi identities if we define the exterior covariant
derivative of a $\g$-valued differential form $X$ by
\[    d_A X = dX + A \we X , \]
and of an $\h$-valued differential form $Y$ by 
\[    d_A Y = dY + \underline{A \we Y} . \]
It is then well-known that 
\[    d_A^2 X = F \we X , \qquad d_A^2 Y = \underline{F \we Y}.\]
In this notation we have
\[       G = d_A B . \]

\begin{prop} \et If $(A,B)$ is any connection on $M$, 
then its curvature $(F,G)$ satisfies the {\bf higher Bianchi identities}:
\begin{equation} \label{bianchi} 
\begin{array}{lcl} 
       d_A F &=& -\underline{G}  \\
       d_A G &=& \underline{F \we B}  . \\
\end{array}
\end{equation}
\end{prop}

Proof --- For the first equation, note that
\[
\begin{array}{lcl}
d_A F &=& dF + A \we F \\
      &=& d(dA + \hf A \we A - \underline{B}) + 
          A \we(dA + \hf A \we A - \underline{B}) \\
      &=& - d_A \underline{B} \\
      &=& - \underline{d_A B} = - \underline G
\end{array}
\]
where we use the fact that
for any $\h$-valued form $Y$ we have
\[  d_A \underline{Y} = \underline{d_A Y}  \]
thanks to equation (\ref{cross1}).   For the second equation,
note that $d_A G = d_A^2 B = \underline{F \we B}$. \qed

\noindent
As we shall see, the higher Yang--Mills equations have a very
similar appearance to the higher Bianchi identities.

\section{Higher Yang--Mills Equations} \label{ymeqs}

To set up Yang--Mills theory one assumes the spacetime manifold
is equipped with a metric, and for convenience usually an orientation
as well.  Also, one needs the gauge group to have a nondegenerate invariant 
symmetric bilinear form on its Lie algebra.  Here we wish to set
up a version of `higher Yang--Mills theory', so we will need similar 
extra structure on our manifold $M$ and our Lie 2-group $C$.

In particular, suppose that $M$ is an oriented manifold equipped with a
semi-Riemannian metric. Suppose also that $C$ is a Lie 2-group whose
differential crossed module $(\g,\h,dt,d\alpha)$ has the following extra
structure: $\g$ is equipped with a nondegenerate symmetric bilinear form
that is invariant under the adjoint action of $G$, and $\h$ is equipped
with a nondegenerate symmetric bilinear form that is invariant under
the adjoint action of $H$ and also the action of $G$ coming from $\alpha
\maps G \to \Aut(H)$.  If we write both bilinear forms as $\langle
\cdot, \cdot \rangle$, these invariance conditions imply that
\begin{equation}\label{inner.1}
     \langle [x,x'],x'' \rangle = -\langle x', [x,x''] \rangle , 
\end{equation}
\begin{equation}\label{inner.2}
     \langle [y,y'],y'' \rangle = -\langle y', [y,y''] \rangle , 
\end{equation}
and
\begin{equation}\label{inner.3}
    \langle [x,y],y' \rangle = -\langle y, [x,y'] \rangle 
\end{equation}
for all $x,x',x'' \in \g$ and $y,y',y'' \in \h$.  Note that in the last
equation we are using the bracket $[x,y]$ of $x \in \g$ and $y \in \h$
to stand for $d\alpha(x)(y)$.

Copying the usual formula for the Yang--Mills action, we write down 
the following action as a function of the connection $(A,B)$:
\begin{equation} \label{ym.action}
   S(A,B) = \int_M \tr(F \we \ast F) + \tr(G \we \ast G)  .
\end{equation}
Here, as usual, we define
\[       \tr((x \tensor \mu) \we (x' \tensor \nu)) = \langle x,y\rangle \; 
\mu \we \nu \]
whenever $x,x' \in \g$ and $\mu,\nu$ are differential forms on $M$;
this extends by linearity in each argument to all $\g$-valued differential 
forms, and we use the same sort of formula for $\h$-valued differential
forms.  Also as usual, we define the Hodge dual of a $\g$-valued
differential form by
\[    \ast (x \tensor \mu) = x \tensor \ast \mu  \]
and similarly for $\h$-valued differential forms.  

Starting with this action we can obtain the equations of motion
by taking the variational derivative $\delta S$ and setting
$\delta S = 0$.    To prepare for the calculation note that
\[        \delta F = d_A \delta A - \underline{\delta B}  \]
and 
\[        \delta G = d_A \delta B + \underline{\delta A \we B}   .\]
We thus have
\[
\begin{array}{lcl}
\delta S &=& 2 \int_M \tr(\delta F \we F) + \tr (\delta G \we \ast G) \\
&& \\
&=& 2 \int_M \tr\left((d_A \delta A - \underline{\delta B}) \we \ast F\right) 
+ \tr\left((d_A \delta B + \underline{\delta A \we B}) \we \ast G\right) \\
&& \\
&=& 2 \int_M \tr(\delta A \we d_A \ast F - \underline{\delta B} \we \ast F) 
+ \tr(-\delta B \we d_A \ast G + (\underline{\delta A \we B}) \we \ast G) \\
\end{array}
\]
where in the last step we do integration by parts with the help of
equations (\ref{inner.1}) and (\ref{inner.2}).  We ignore boundary terms by
assuming the variations $\delta A$ and $\delta B$ are compactly
supported.

To go further with the last term in the above expression, note that
there is a unique linear map
\[  \sigma \maps \h \tensor \h \to \g   \]
such that
\[       \langle x, \sigma(y,y') \rangle  =
         \langle [x,y], y' \rangle  \]
for all $x \in \g$ and $y,y' \in \h$.  Given 
$\h$-valued differential forms $Y = y \tensor \nu$ and $Y' = 
y' \tensor \nu'$, we define
\[      \underline{Y \we Y'} = \sigma(y,y') \tensor (\nu \we \nu'), \]
and we extend this to arbitrary $\h$-valued differential forms by 
linearity.  Using this notation it follows that
\[  \delta S = 
 2 \int_M \tr(\delta A \we d_A \ast F - \underline{\delta B} \we \ast F) + 
\tr(-\delta B \we d_A \ast G + \delta A \we (\underline{B \we \ast G}))  .\]
Furthermore, note that the map $\sigma$ is skew-symmetric,
since
\[       \langle x, \sigma(y',y) \rangle  = 
       \langle [x,y'], y \rangle  = -\langle y', [x,y] \rangle
      = \langle x, \sigma(y,y') \rangle \]
for all $x \in \g$ and $y,y' \in \h$, by equation (\ref{inner.3}).
It follows that 
\[      \underline{B \we \ast G} = - \underline{\ast G \we B} ,\]
and it will be convenient to use this to rewrite the above formula as
\[  \delta S = 
 2 \int_M \tr(\delta A \we d_A \ast F - \underline{\delta B} \we \ast F) -
\tr(\delta B \we d_A \ast G + \delta A \we (\underline{\ast G \we B}))  .\]

Next we rewrite the second term in $\delta S$ by defining a map
\[        dt^* \maps \h \to \g \]
such that
\[     \langle x, dt^*(y) \rangle = \langle dt(x) , y \rangle \]
for all $x \in \g$, $y \in \h$.  Given a $\g$-valued differential form
$X = x \tensor \mu$, we define
\[       \underline{X} = dt^*(x) \tensor \mu \]
and extend this to arbitrary $\g$-valued differential forms by 
linearity.  Using this we obtain
\[
\begin{array}{lcl} 
\nonumber  \delta S &=&
 2 \int_M \tr(\delta A \we d_A \ast F - \delta B \we \ast \underline{F}) -
\tr(\delta B \we d_A \ast G + \delta A \we (\underline{\ast G \we B}))  \\
&& \\
&=& 
 2 \int_M \tr\left(\delta A \we (d_A \ast F - \underline{\ast G \we B})\right) 
- \tr\left(\delta B \we (d_A \ast G + \ast \underline{F})\right)  .
\end{array}
\]
From this we see:

\begin{thm} \label{YM} \et Suppose that $M$ is an oriented
semi-Riemannian manifold, and suppose that the Lie 2-group $C$ has an
associated Lie crossed module $(G,H,t,\alpha)$ for which $\g$ and $\h$
are equipped with nondegenerate invariant symmetric bilinear forms.
Then the connection $(A,B)$ on $M$ satisfies $\delta S = 0$ if and
only if the {\bf higher Yang--Mills equations} hold:
\begin{equation} \label{ym.equations}
\begin{array}{lcl} 
d_A \ast F &=& \underline{\ast G \we B}   \\
d_A \ast G &=& - \ast \underline{F}  
\end{array}
\end{equation}
\end{thm}

Collecting the higher Bianchi identities and higher Yang--Mills
equations together, we have:
\[
\begin{array}{lclrlcl}
d_A F &=& - \underline{G} &\hbox{\hskip 3em} &
d_A \ast G &=&  - \ast \underline{F} \\
d_A G &=&  \underline{F \we B} &&
d_A \ast F &=& \underline{\ast G \we B}   \\
\end{array}
\]
These equations have a pleasant symmetry.  In particular,
the Bianchi identities would imply the Yang--Mills equations if we
had
\[      \ast F = G, \qquad  \ast G = F .\]
These conditions are reminiscent of the self-dual Yang--Mills equations
on a 4-dimensional Riemannian manifold.  However, they only make sense
under certain conditions.  First, since $F$ is a 2-form and $G$ is a
3-form, we must be in 5 dimensions.  Second, to obtain $\ast^2 = 1$ on
2-forms in 5-dimensional space we need the metric to have a signature
with an even number of minus signs.  Third, since $F$ is $\g$-valued and
$G$ is $\h$-valued, we need an isomorphism $\g \iso \h$ in terms of which
all the underline maps become identity maps.  One way to achieve this is to
let $H$ be a semisimple Lie group and let $C$ be its automorphism
2-group, as described in Example \ref{crossed.5}.  In this case
we have:

\begin{thm} \label{selfdual} \et  
Suppose that $M$ is a 5-dimensional oriented semi-Riemannian manifold of
signature $(+++++)$, $(+++--)$ or $(+----)$.  Let $H$ be a semisimple
Lie group, let $C$ be the automorphism 2-group of $H$, and let
$(G,H,t,\alpha)$ be the associated Lie crossed module.  If we define the
higher Yang--Mills action using the Killing forms on $\g$ and $\h$, then
the connection $(A,B)$ on $M$ is a solution of the higher Yang--Mills
equations if $\ast F = \underline{G}$. 
\end{thm}

Proof --- Since $H$ is semisimple, the Killing form on $\h$
is a nondegenerate symmetric bilinear form that is invariant
under the adjoint action of $H$.  Moreover, the connected component 
of $G = \Aut(H)$ consists of inner automorphisms, so it is isomorphic
to $H$ modulo its center, which is discrete.  This gives an isomorphism 
$\g \iso \h$.  Using this isomorphism to identify $\g$ with $\h$,
one can check that all the underline maps act trivially.  
The higher Bianchi identities thus become simply
\[
\begin{array}{lcl} 
       d_A F &=& -G \\
       d_A G &=& F \we B   \\
\end{array}
\]
and these imply the higher Yang--Mills equations
\[
\begin{array}{lcl} 
d_A \ast G &=& - \ast F  \\
d_A \ast F &=& \ast G \we B   \\
\end{array}
\]
if we assume $\ast F = G$, and thus $\ast G = F$.  \qed

It would be interesting to study the moduli space of self-dual solutions
of the higher Yang--Mills equations in 5 dimensions, and see if they
have any implications for topology.  Of course this involves the concept
of `gauge transformation' for connections on 2-bundles. 

\subsection*{Acknowledgements}  I thank Lawrence Breen for very
helpful discussions and correspondence.

\end{document}